\documentclass[unnumsec,webpdf,modern,large]{oup-authoring-template}

\usepackage{booktabs}
\usepackage{makecell}
\usepackage{xcolor}
\usepackage{colortbl}
\usepackage{soul}
\usepackage{cleveref}

\graphicspath{{figs/}}

\theoremstyle{thmstyleone}%
\theoremstyle{thmstyletwo}%
\theoremstyle{thmstylethree}%

\newcommand{\rev}[1]{\textcolor{black}{#1}}
\newcommand{\revbegin}{\color{black}}
\newcommand{\revend}{\color{black}}

\begin{document}

\journaltitle{Journal Title Here}
\DOI{DOI HERE}
\copyrightyear{2021}
\pubyear{2021}
\access{Advance Access Publication Date: Day Month Year}
\appnotes{Preprint}

\firstpage{1}


\title[Cross study analysis of drug response prediction]{A cross-study analysis of drug response prediction in cancer cell lines}

\author[1,$\ast$]{Fangfang Xia}
\author[2]{Jonathan Allen}
\author[1]{Prasanna Balaprakash}
\author[1]{Thomas Brettin}
\author[3]{Cristina Garcia-Cardona}
\author[1,5]{Austin Clyde}
\author[3]{Judith Cohn}
\author[4]{James Doroshow}
\author[5]{Xiaotian Duan}
\author[6]{Veronika Dubinkina}
\author[7]{Yvonne Evrard}
\author[2]{Ya Ju Fan}
\author[3]{Jason Gans}
\author[2]{Stewart He}
\author[7]{Pinyi Lu}
\author[6]{Sergei Maslov}
\author[1]{Alexander Partin}
\author[1]{Maulik Shukla}
\author[7]{Eric Stahlberg}
\author[1]{Justin M. Wozniak}
\author[1]{Hyunseung Yoo}
\author[7]{George Zaki}
\author[1]{Yitan Zhu}
\author[1,5]{Rick Stevens}

\authormark{Xia et al.}

\address[1]{Argonne National Laboratory}
\address[2]{Lawrence Livermore National Laboratory}
\address[3]{Los Alamos National Laboratory}
\address[4]{National Cancer Institute}
\address[5]{University of Chicago}
\address[6]{University of Illinois at Urbana-Champaign}
\address[7]{Frederick National Laboratory for Cancer Research}


\corresp[$\ast$]{Corresponding author. \href{email:fangfang@anl.gov}{fangfang@anl.gov}}

\received{Date}{0}{Year}
\revised{Date}{0}{Year}
\accepted{Date}{0}{Year}



\abstract{
To enable personalized cancer treatment, machine learning models have been developed to predict drug response as a function of tumor and drug features.
However, most algorithm development efforts have relied on cross validation within a single study to assess model accuracy.
While an essential first step, cross validation within a biological data set typically provides an overly optimistic estimate of the prediction performance on independent test sets.
To provide a more rigorous assessment of model generalizability between different studies, we use machine learning to analyze five publicly available cell line-based data sets: NCI60, CTRP, GDSC, CCLE and gCSI.
Based on observed experimental variability across studies, we explore estimates of prediction upper bounds.
We report performance results of a variety of machine learning models, with a multitasking deep neural network achieving the best cross-study generalizability.
By multiple measures, models trained on CTRP yield the most accurate predictions on the remaining testing data, and gCSI is the most predictable among the cell line data sets included in this study.
With these experiments and further simulations on partial data, two lessons emerge: (1) differences in viability assays can limit model generalizability across studies, and (2) drug diversity, more than tumor diversity, is crucial for raising model generalizability in preclinical screening.
}



\maketitle

\begin{table*}
 \caption{Characteristics of drug response data sets included in the cross-study analysis}
  \label{tab:datasets}
  \centering
\begin{tabular*}{\textwidth}{@{\extracolsep{\fill}}lrrrrl@{\extracolsep{\fill}}}
    \toprule
    Data Set & Cells & Drugs & Dose Response Samples & Drug Response Groups$^{1}$ & Viability Assay \\
    \midrule
    NCI60 & 60 & 52,671 & 18,862,308 & 3,780,148 & Sulforhodamine B stain \\
    CTRP & 887 & 544 & 6,171,005 & 395,263 & CellTiter Glo \\
    CCLE & 504 & 24 & 93,251 & 11,670 & CellTiter Glo \\
    GDSC & 1,075 & 249 & 1,894,212 & 225,480 & Syto60 \\
    gCSI & 409 & 16 & 58,094 & 6,455 & CellTiter Glo \\
    \bottomrule
  \end{tabular*}
  \begin{tablenotes}%
    \item[$^{1}$] We define a response group as the set of dose response samples corresponding to a particular pair of drug and cell line.
  \end{tablenotes}
\end{table*}

\section{Introduction}
\label{sec:introduction}
Precision oncology aims at delivering treatments tailored to the specific characteristics of a patient's tumor.
This goal is premised on the idea that, with more data and better computational models, we will be able to predict drug response with increasing accuracy.
Indeed, two recent trends support this premise.
First, high-throughput technologies have dramatically increased the amount of pharmacogenomic data.
A multitude of omic types such as gene expression and mutation profiles can now be examined for potential drug response predictors.
On the prediction target side, while clinical outcomes are still limited, screening results abound in preclinical models that mimic patient drug response with varying fidelity.
Second, deep learning has emerged as a natural technique to capitalize on the data.
Compared with traditional statistical models, the high capacity of neural networks enable them to better capture the complex interactions among molecular and drug features.

The confluence of these factors has ushered in a new generation of computational models for predicting drug response.
In the following, we provide a brief summary of recent work in this field, with a focus on representative studies using cancer cell line data.

The NCI60 human tumor cell line database \cite{shoemaker2006nci60} was one of the earliest resources for anticancer drug screen.
Its rich molecular characterization data have allowed researchers to compare the predictive power of different assays.
For example, gene expression, protein, and microRNA abundance have been shown to be effective feature types for predicting both single and paired drug response \cite{cortes2016improved,xia2018predicting}.
In the last decade, new resources including Cancer Cell Line Encyclopedia (CCLE) \cite{barretina2012cancer,ghandi2019next}, Genomics of Drug Sensitivity in Cancer (GDSC) \cite{yang2012genomics}, and Cancer Therapeutics Response Portal (CTRP) \cite{basu2013interactive,seashore2015harnessing}, have significantly expanded the number of screened cancer cell lines.
Drug response prediction has moved beyond per-drug or per-cell line analysis \cite{ding2018precision,rampavsek2019dr} to include integrative models that take both drug and cell line as input features.
Community challenges have further inspired computational approaches \cite{menden2019community,douglass2020community}.
A wide range of new machine learning techniques have been explored, including recommendation systems \cite{suphavilai2018predicting}, ranking methods \cite{gerdes2021drug,daoud2020q}, generative models \cite{rampavsek2019dr,kadurin2017cornucopia}, \rev{feature analysis \cite{huang2017open}, network modeling \cite{wei2019comprehensive}}, ensemble models \cite{su2019deep,rahman2019functional,bomane2019paclitaxel,sidorov2019predicting,su2020meta}, and deep learning approaches \cite{zeng2019deepdr,li2019deepdsc,zhang2018novel,chang2018cancer,liu2019improving}, with some incorporating novel design ideas such as attention \cite{oskooei2018paccmann} and visual representation of genomic features  \cite{bazgir2020representation}.
A number of excellent review articles have recently been published on the topic of drug response prediction, with substantial overlap and special emphases on data integration \cite{li2018review}, feature selection \cite{ali2019machine}, experimental comparison \cite{pucher2019comparison}, machine learning methods \cite{adam2020machine}, systematic benchmarking \cite{chen2021survey}, combination therapy \cite{wang2020machine}, deep learning results \cite{baptista2021deep}, and meta-review \cite{guvencc2021improving}.

Despite the tremendous progress in drug response prediction, significant challenges remain:
(1) Inconsistencies across studies in genomic and response profiling have long been documented \cite{haibe2013inconsistency,mpindi2016consistency,safikhani2016revisiting,sadacca2017new}.
The Genentech Cell Line Screening Initiative (gCSI)  \cite{haverty2016reproducible} was specifically designed to investigate the discordance between CCLE and GDSC by an independent party.
While harmonization practices help \cite{smirnov2017pharmacodb,rahman2019evaluating,gupta2020normalized}, a significant level of data variability will likely remain in the foreseeable future due to the complexity in cell subtypes and experimental standardization.
(2) The cross-study data inconsistencies suggest that a predictive model trained on one data source may not perform as well on another.
Yet, most algorithm development efforts have relied on cross validations within a single study, which likely overestimate the prediction performance.
Even within a single study, validation $R^2$ (explained variance) rarely exceeds $0.8$ when strict data partition is used \cite{baptista2021deep}, indicating difficulty in model generalization.
(3) Without a single study that is sufficiently large, a natural next step is to pool multiple data sets to learn a joint model.
Multiple efforts have started in this direction, although the gains to date from transfer learning or combined learning have been modest \cite{dhruba2018application,zhu2020ensemble,clyde2020systematic}.
(4) It is also unclear how model generalization improves with increasing amounts of cell line or drug data.
This information will be essential for future studies to prioritize screening experiments.

In this study, we seek to provide a rigorous assessment of the performance range for drug response prediction given the current publicly available data sets.
Focusing on five cell line screening studies,
we first estimate within- and cross-study prediction upper bounds based on observed response variability.
After integrating different drug response metrics, we then apply machine learning to quantify how models trained on a given data source generalize to others.
These experiments provide insights on the data sets that are more predictable or contain more predictive value.
To understand the value of new experimental data, we further use simulations to study the relative importance of cell line versus drug diversity and how their addition may impact model generalizability.

\section{Data integration}

Pan-cancer screening studies have been comprehensively reviewed by Baptista et al. \cite{baptista2021deep}.
In this study, we focus on single-drug response prediction and include five public data sets: NCI60, CTRP, GDSC, CCLE, and gCSI.
The characteristics of the drug response portion of these studies are summarized in Table \ref{tab:datasets}.
These data sets are among the largest in response sample size of the nine data sources reviewed and, therefore, have been frequently used by machine learning studies.
Together, they also capture a good representation of the diversity in viability assay for response profiling.

\subsection{Data acquisition and selection}
Of the five drug response data sets included in this study, NCI60 was downloaded directly from the NCI FTP, and the remaining four were from PharmacoDB \cite{smirnov2017pharmacodb}.
We use GDSC and CTRP to denote the GDSC1000 and CTRPv2 data collections in PharmacoDB, respectively.

The different experimental designs of these studies resulted in differential coverage of cell lines and drugs.
While NCI60 screened the largest compound library on a limited number of cell lines, CCLE and gCSI focused on a small library of established drugs with wider cell line panels.
The remaining two studies, CTRP and GDSC, had more even coverage of hundreds of cell lines and drugs.

There is considerable \rev{overlap} in cell lines and drugs covered by the five data sets.
A naive partitioning of the samples by identifier would thus lead to leakage of training data into the test set by different aliases.
Thanks to PharmacoDB's curation and integration effort, such alias relationships can be clearly identified through name mapping.

To create a working collection of samples, we first defined a cell line set and a drug set.
Our cell line selection is the union of all cell lines covered by the five studies.
For drugs, however, NCI60 screened a collection of over 50,000 drugs that would be too computationally expensive for iterative analysis.
Therefore, we selected a subset of 1,006 drugs from NCI60, favoring FDA-approved drugs and those representative of diverse drug classes.
From the remaining four studies, we included all the drugs.
This resulted in a combined set of 1,801 drugs.

\subsection{Drug response integration}
For cell line-based anticancer screening, drug sensitivity is quantified by dose response values that measure the ratio of surviving treated to untreated cells after exposure to a drug treatment at a given concentration.
The dose response values can be further summarized, for a particular cell-drug pair, to form dose-independent metrics of drug response.

All five studies provided dose response data, but with different ranges.
To facilitate comparison across studies, dose response values were linearly rescaled to share a common range from $-100$ to $100$.

Dose-independent metrics were more complicated.
For example, whereas GDSC used $IC_{50}$ (dose of $50\%$ inhibition of cell viability), NCI60 used $GI_{50}$ (dose at which $50\%$ of the maximum growth inhibition is observed).
Given the lack of consensus, we adopted PharmacoDB's practice to recompute the aggregated dose-independent metrics for each study from the individual dose response values, removing inconsistencies in calculation and curve fitting \cite{smirnov2017pharmacodb}.

\begin{figure*}
  \centering
  \includegraphics[width=0.95\linewidth]{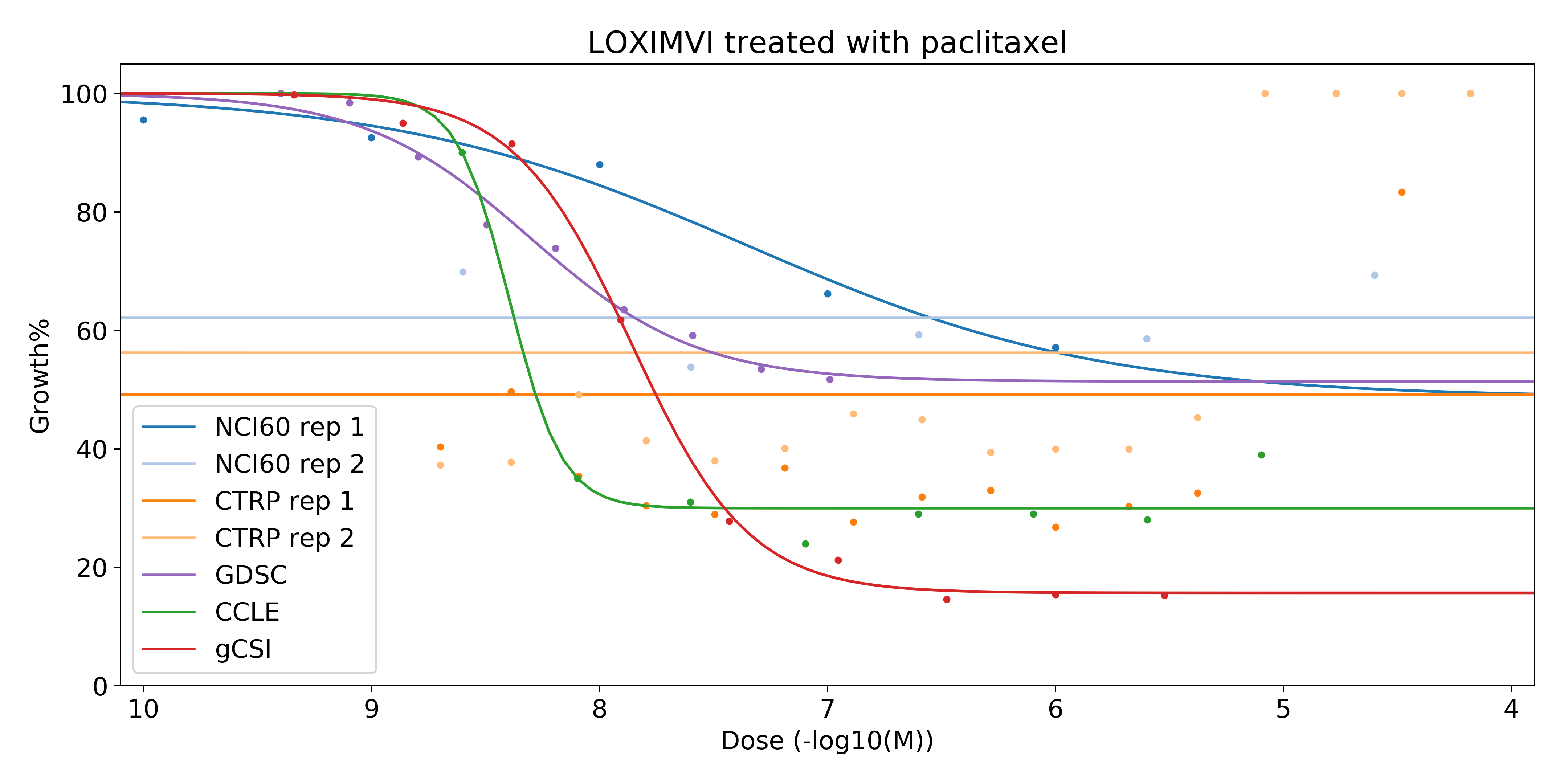}
  \caption{{\bf Fitted dose response curves from multiple studies.} This example shows the dose response of the LOXIMVI melanoma cell line treated with paclitaxel. Curves have been consistently fitted across the studies. Experimental measurements from multiple sources and replicates are not in complete agreement.}
  \label{fig:curves}
\end{figure*}

Among these summary statistics, AAC is the area above the dose response curve for the tested drug concentrations, and DSS (drug sensitivity score) is a more robust metric normalized against dose range \cite{yadav2014quantitative}.
The definitions of these two metrics were directly borrowed from PharmacoDB, but they had to be recomputed in this study because PharmacoDB does not cover NCI60.
We used the same Hill Slope model with three bounded parameters \cite{smirnov2017pharmacodb} to fit the dose response curves.
We also added another dose-independent metric, AUC, to capture the area under the curve for a fixed concentration range ($[10^{-10}, 10^{-4}]{\mu}M$).
AUC can be viewed as the average growth and can be compared across studies.
\revbegin
See the Methods section for the detailed definitions of these metrics.
\revend

\section{Results}

\subsection{Response prediction upper bounds}
\revbegin
Data sets from different studies often exhibit biases.
Apart from differences in study design (e.g., choice of cell lines and drugs, sample size), there are two main types of biases associated with experimentally measured data: bias in molecular characterization, and bias in response assay.
Viewed in the context of a machine learning problem, these correspond to biases in input features and output labels.
They form the main challenge to joint learning and transfer learning efforts over multiple data sources.

The first type of bias can be alleviated with feature preprocessing.
Specifically, we used batch correction to homogenize gene expression profiles from different databases and followed a consistent protocol to generate drug features and fingerprints (see our previous work \cite{clyde2020systematic} and a summary in the Methods section).
\revend

\rev{The second type of bias is the focus of this paper.}
\rev{That is,} despite the integration efforts, significant differences in measured drug response remain.
Fig. \ref{fig:curves} illustrates this heterogeneity \rev{with a common example: the LOXIMVI melanoma cell line treated with Paclitaxel.
This same combination is included by multiple studies, but it exhibits distinct response curves across replicates as well as studies}.
Given the degree of variability in drug response measurements, it is natural to ask what's the best prediction models could do with infinite data.
We explored this question based on within- and cross-study replicates.

\subsubsection{Within-study response variability among replicates}
Three of the studies (NCI60, CTRP, GDSC) contained replicate samples where the same pair of drug and cell line were assayed multiple times.
This data allowed us to estimate within-study response variability as summarized in Table \ref{tab:var_within}.
The fact that these studies used three different viability assays was also an opportunity to sample the associations between variability level and assay technology.

\begin{table*}
  \setlength{\tabcolsep}{9pt}
 \caption{Dose response variability among replicates in the same study}
 \centering
 \begin{tabular}{p{1cm}p{1.8cm}p{1.8cm}p{2.3cm}p{3.5cm}p{2.9cm}}
    \toprule
    Study & Samples with Replicates & Replicates per Group & Mean Response S.D. in Group & R$^2$ Explaining Response with Group Mean & R$^2$ for Samples with Replicates \\
    \midrule
    NCI60 & 41.56\% & 2.62 & 14.5 & 0.959 & 0.931 \\
    CTRP & 4.09\% & 2.05 & 18.8 & 0.996 & 0.862 \\
    GDSC & 2.62\% & 2.00 & 21.9 & 0.996 & 0.810 \\
    \bottomrule
  \end{tabular}
  \label{tab:var_within}
\end{table*}\

Overall, NCI60 had the lowest standard deviation in dose response at\rev{$14.5$} for an average replicate group after rescaling response values linearly in all three studies to a common range.
To establish a ceiling for the performance of machine learning models, we computed how well the mean dose response of a replicate group would predict individual response values.
This resulted in an $R^2$ score of $0.959$ for NCI60.
The equivalent values for CTRP and GDSC were much higher, but only for the fact that they had much lower fractions of samples with replicates.
When we confined the analysis to just non-singleton replicate groups, the $R^2$ scores dropped significantly with GDSC being the lowest at $0.81$.
While it's unclear how well these replicate groups represent the entire study, we should expect machine learning models to not exceed the $R^2$ score in cross validation.
\rev{This is not a statement on the capability of the machine learning methods, but the noise level in the ground truth data that they work with.}

\subsubsection{Cross-study response variability}
For comparing drug response across different studies, dose-dependent metrics are less useful, because each study generally has its own experimental concentrations.
Instead, we opted for the three aforementioned dose-independent response metrics: AUC, AAC, and DSS.
A key difference between AUC and AAC is that AUC is based on a fixed dose range while AAC is calculated for the drug-specific dose window screened in a particular study.
DSS is another metric that integrates both drug efficacy and potency, with evidence of more differentiating power in understanding drug functional groups \cite{yadav2014quantitative}.

Using these three metrics, we analyzed cross-study response
\revbegin
variability by focusing on the common subset of cell line and drug pairs that appear in multiple studies.
\revend
Table \ref{tab:var_cross} summarizes two representative cases.
In the first scenario,  we used response values from CTRP to directly predict the corresponding metrics in CCLE.
In this case, the source and target studies shared a common viability assay.
Yet, the raw $R^2$ on all three metrics were only around $0.6$, with AAC being the highest.
In the second scenario, where the target study GDSC used a different viability assay, the response projections were markedly worse: the best $R^2$ score was slightly above $0.3$ achieved on AUC, and the other two scores were close to zero.

\begin{figure*}
    \centering
    \subfloat{{\includegraphics[width=0.45\linewidth]{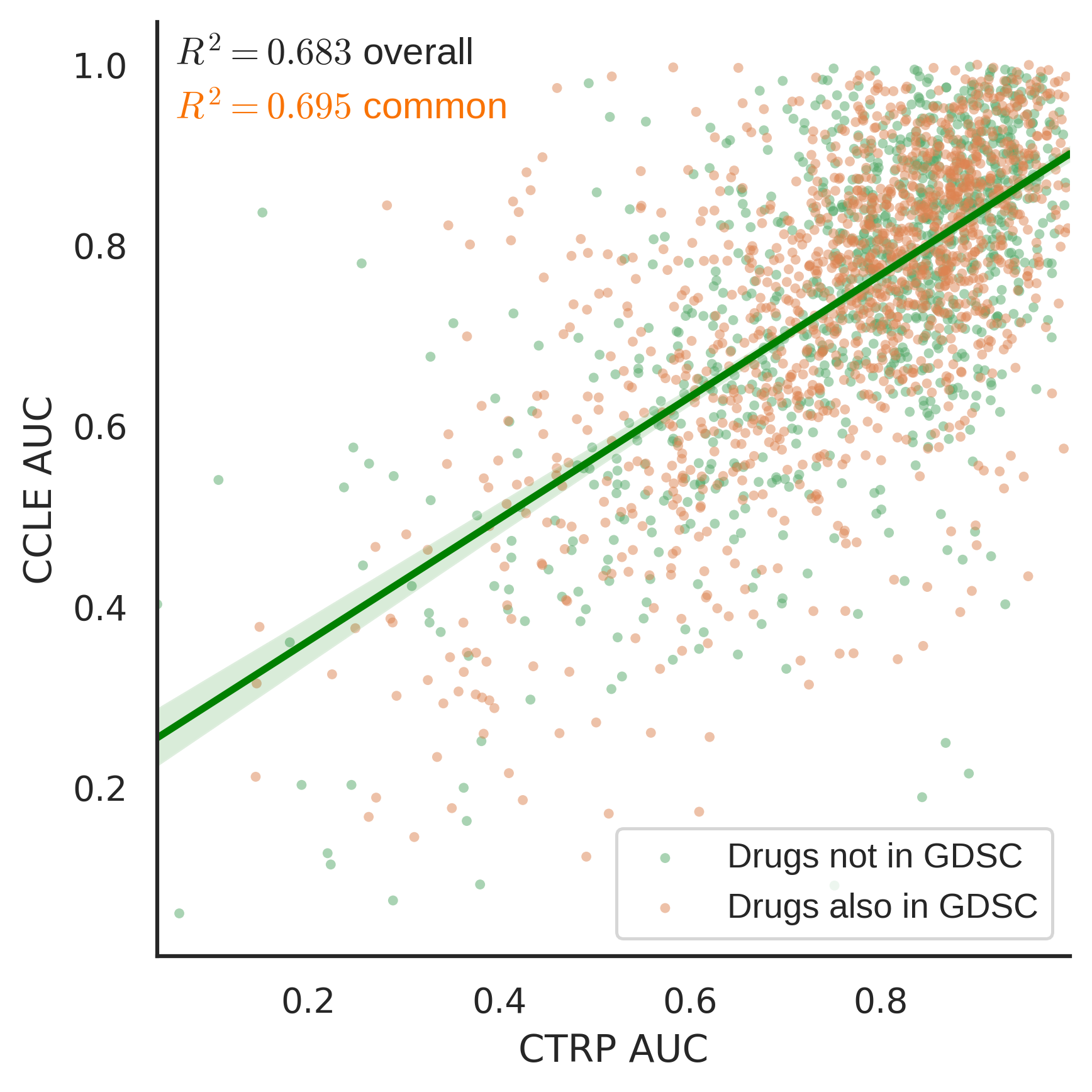} }}%
    \qquad
    \subfloat{{\includegraphics[width=0.45\linewidth]{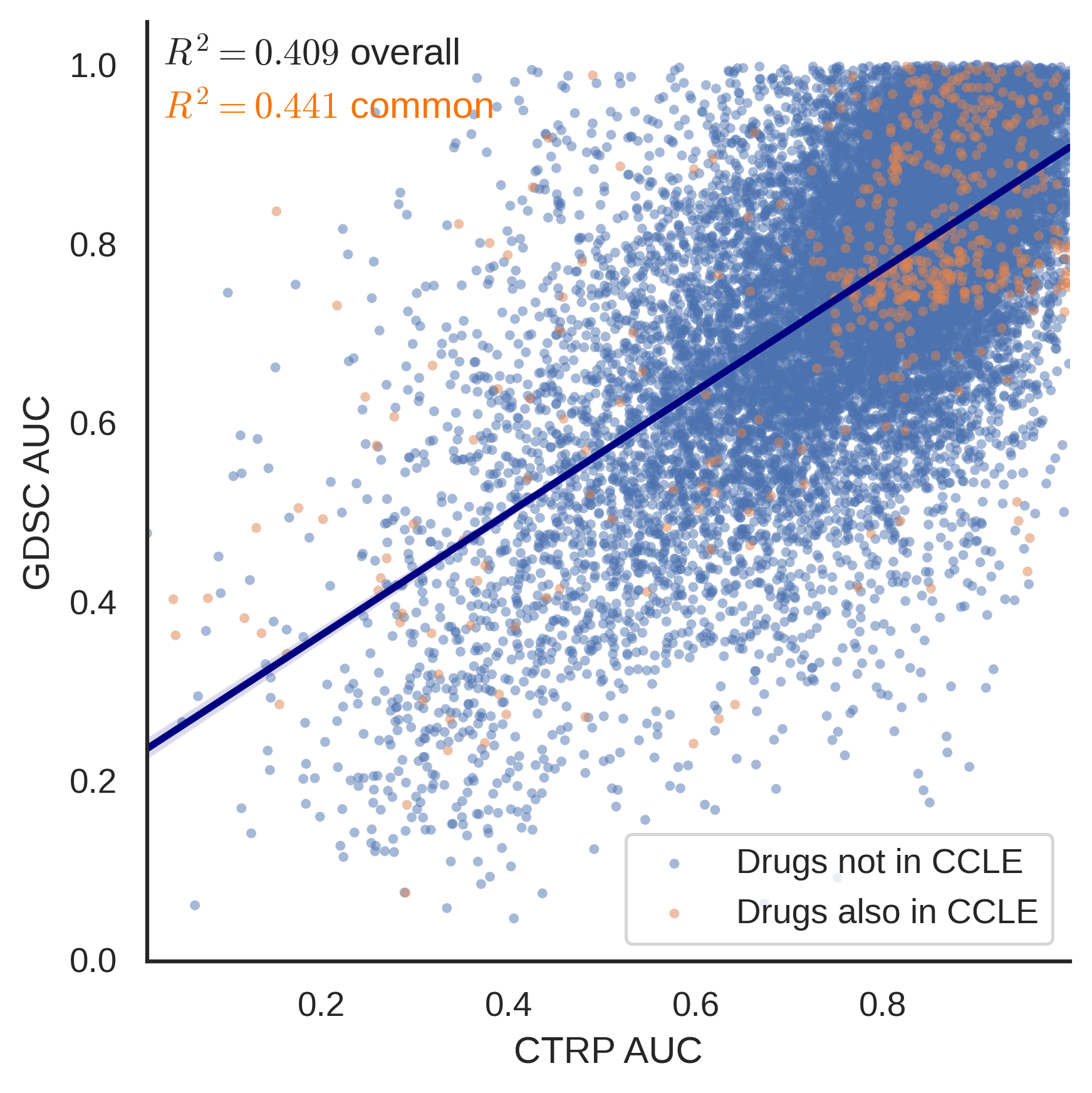} }}%
    \caption{\rev{{\bf Estimating cross-study response variability based on overlapping experimental data.}
      When the same combination of drug and cell line appears in multiple studies, we can use the reported differences to estimate cross-study response variability.
      Here we map the AUC (area under drug response curve) values from CTRP to CCLE (left) and CTRP (right) in the scatter plots with linear regression fit.
      Orange dots represent experiments involving common drugs shared by CTRP, CCLE, and GDSC, reducing sampling bias in different studies.
      In both plots, $R^2$ scores are reported separately for the overall fit and the subset with common drugs.
      Overall, there is greater agreement between CTRP and CCLE (same assay) than between CTRP and GDSC (different assay).}
    }
    \label{fig:var}
\end{figure*}

\revbegin
In the above analysis, the response values from one source were straightly used as the predicted response in another source.
This is applicable to cases where the response information in the target study is unknown.
When the target distribution is known, a mapping function can be used to fit the response values, as a way to bridge the measurement differences between the studies.
Fig. \ref{fig:var} illustrates this with simple linear regression.
This improved the best $R^2$ scores (see the \emph{Fit} columns in Table \ref{tab:var_cross}) to $0.68$ for the mapping from CTRP to CCLE (same assay), and $0.41$ going from CTRP to GDSC (different assays).
\revend
Again, these numbers only gave rough estimates based on partial data; they nevertheless calibrated what we could expect from machine learning models performing cross-study prediction.

\revbegin
So far we have shown that there is greater variability between CTRP and GDSC than between CTRP and CCLE, but it's unclear whether this can be attributed to the difference in response assay alone.
Another difference in drug diversity between the target studies is at play, with GDSC having far more drugs (249) than CCLE (24).
To tease apart these two factors, we narrowed down to a subset of seven drugs shared by all three studies: Paclitaxel, Lapatinib, AZD0530, AZD6244, Nilotinib, Erlotinib, and Sorafenib.
Scatter plots of response data points involving this common drug set are highlighted in orange in Fig. \ref{fig:var}.
Linear regression fit on the subset resulted in slightly higher $R^2$ scores: $0.69$ between CTRP and CCLE, and $0.44$ between CTRP and GDSC, both on the dose-independent AUC metric.
Their unchanging relationship suggests that the observed cross-study response variability is more predominantly due to viability assay than sampling difference.

This comparison between CellTiter Glo (CTRP, CCLE) and Syto60 (GDSC) does not necessarily generalize to other pairs of viability assays.
NCI60 is another study that used a different viability assay.
When we swapped GDSC for NCI60 in the above analysis, the overall $R^2$ score from CTRP to NCI60 is 0.67, only slightly lower than that to CCLE.
The $R^2$ score on the common set among CTRP, NCI60, and CCLE is lower at 0.60, suggesting that sampled drug diversity also plays a role in the estimate of cross-study response variability.
\revend

\captionsetup[subfigure]{position=top,labelformat=empty}

\begin{table*}
\begin{center}
 \caption{Dose-independent response variability in identical cell-drug pairs across studies}
 \label{tab:var_cross}
  \begin{tabular*}{\textwidth}{@{\extracolsep{\fill}}ccrrrrrrr@{\extracolsep{\fill}}}
    \toprule
    \multirow{1}{*}{Source} &
    \multirow{1}{*}{Target} &
    \multicolumn{1}{c}{Overlapping} &
    \multicolumn{2}{c}{R$^2$ on AUC$^{1}$} &
    \multicolumn{2}{c}{R$^2$ on AAC} &
    \multicolumn{2}{c}{R$^2$ on DSS$^{2}$} \\
    \cline{4-5} \cline{6-7} \cline{8-9}
    Study & Study & Cell-Drug Groups & Raw$^{3}$ & \rev{Fit$^{4}$} & Raw & \rev{Fit} & Raw & \rev{Fit} \\
    \midrule
    CTRP & CCLE &  2,339 & 0.594 & \rev{0.683} & 0.641 & \rev{0.681} & 0.635 & \rev{0.670} \\
    CTRP & GDSC & 17,259 & 0.302 & \rev{0.409} & 0.019 & \rev{0.197} & 0.006 & \rev{0.194} \\
    \bottomrule
  \end{tabular*}
  \begin{tablenotes}%
    \item[$^{1}$] AUC (area under drug response curve) is not a direct complement of AAC (area above drug response curve). They are defined on different concentration ranges: AUC for a fixed range ($[10^{-10}, 10^{-4}]{\mu}M$), and AAC for study-specific, tested concentration ranges.
    \item[$^{2}$] DSS (drug sensitivity score) is a normalized variation of AAC defined by Yadav et al. \cite{yadav2014quantitative}.
    \rev{
    \item[$^{3}$] Raw R$^2$ scores (explained variance) were derived by directly comparing the corresponding drug response values, for a given metric, on the common experiments with cell-drug pairs shared by the two studies (see scatter plots in Fig \ref{fig:var}).
    \item[$^{4}$] To account for study bias in experimental response measurement, linear regression was used to fit the response values on the same cell-drug pair between source and target studies (see regression lines in Fig \ref{fig:var}).
    }
  \end{tablenotes}
\end{center}
\end{table*}


\subsection{Cross-study prediction of drug response}

We applied machine learning models to predict dose response.
When evaluated in the context of a single data set, the prediction performance is dependent on two general factors: the input features and the model complexity.
The models we reviewed in the introduction section have offered wide-ranging configurations on these two factors.
Most of these models, however, stopped at evaluation by cross-validation within a single study.
Given the cross-study variability observed, such evaluation likely overestimated the utility of drug response models for practical scenarios involving more than one data sets.

To test how well models trained on one data set could generalize to another, we went beyond study boundaries and tested all combinations of source and target sets.
This introduced a third factor impacting model performance, i.e., the distribution shift between the source and target data sets.
To provide a rigorous assessment of cross-study prediction performance, we applied three machine learning models of increasing complexity.
The first two, Random Forest \cite{breiman2001random} and LightGBM \cite{ke2017lightgbm}, were used as baseline models.
The third one, designed in-house, represented our best effort at a deep learning solution.

\subsubsection{Baseline machine learning performance}
The first baseline performance using Random Forest is shown in Table \ref{tab:RF}.
Each cell in the matrix represents a learning and evaluation experiment involving a source study for training and a target study for testing.
The prediction accuracy for dose response was assessed using both $R^2$ score and mean absolute error (MAE).
Again, the range for drug response values was unified across studies to be from $-100$ to $100$.

\definecolor{green}{rgb}{0.6,0.9,0.6}
\definecolor{yellow}{rgb}{0.9,0.8,0.2}
\definecolor{red}{rgb}{0.9,0.3,0.3}

\newcommand{\mycell}[1]{\small\makecell*{#1}}
\newcommand{\mygcell}[1]{\cellcolor{green!30}\small\makecell*{#1}}
\newcommand{\myycell}[1]{\cellcolor{yellow!25}\small\makecell*{#1}}
\newcommand{\myrcell}[1]{\cellcolor{red!30}\small\makecell*{#1}}
\newcommand{\myGcell}[1]{\cellcolor{green!30}\small\makecell*{\bf#1}}
\newcommand{\myYcell}[1]{\cellcolor{yellow!25}\small\makecell*{\bf#1}}
\newcommand{\myRcell}[1]{\cellcolor{red!30}\small\makecell*{\bf#1}}
\newcommand{\ra}[1]{\renewcommand{\arraystretch}{#1}}

\begin{table*}
  \begin{center}
  \begin{minipage}{0.6\textwidth}
  \caption{Baseline cross-study validation results with Random Forest}
  \label{tab:RF}
  \ra{1.2}
  \begin{tabular}{cccccc}
    \toprule
    \multirow{2}{*}[-10pt]{Training Set} & \multicolumn{5}{c}{Testing Set}\\
    \cmidrule{2-6}
    & NCI60 & CTRP & GDSC & CCLE & gCSI \\
    \midrule
    NCI60 & \mycell{R$^2$ = 0.45\\MAE = 30.4} & \mygcell{R$^2$ = 0.23\\MAE = 34.6} & \mygcell{R$^2$ = 0.15\\MAE = 37.3} & \mygcell{R$^2$ = 0.29\\MAE = 34.3} & \mygcell{R$^2$ = 0.14\\MAE = 54.0} \\
    CTRP & \mygcell{R$^2$ = 0.41\\MAE = 31.7} & \mycell{R$^2$ = 0.30\\MAE = 35.0} & \mygcell{R$^2$ = 0.15\\MAE = 37.4} & \mygcell{R$^2$ = 0.45\\MAE = 29.0} & \mygcell{R$^2$ = 0.17\\MAE = 39.6} \\
    GDSC & \mygcell{R$^2$ = 0.33\\MAE = 36.0} & \mygcell{R$^2$ = 0.14\\MAE = 41.5} & \mycell{R$^2$ = 0.13\\MAE = 40.4} & \mygcell{R$^2$ = 0.17\\MAE = 42.4} & \myycell{R$^2$ = 0.08\\MAE = 43.0} \\
    CCLE & \mygcell{R$^2$ = 0.12\\MAE = 42.6} & \myycell{R$^2$ = -0.03\\MAE = 48.9} & \myrcell{R$^2$ = -0.11\\MAE = 47.1} & \mycell{R$^2$ = 0.17\\MAE = 42.4} & \mygcell{R$^2$ = 0.32\\MAE = 38.5} \\
    gCSI & \myrcell{R$^2$ = -0.38\\MAE = 55.0} & \myrcell{R$^2$ = -0.51\\MAE = 59.0} & \myrcell{R$^2$ = -0.59\\MAE = 58.7} & \myycell{R$^2$ = -0.09\\MAE = 48.6} & \mycell{R$^2$ = 0.25\\MAE = 39.9} \\
    \botrule
  \end{tabular}
  \end{minipage}
  \end{center}
\end{table*}

\begin{table*}
  \begin{center}
  \begin{minipage}{0.6\textwidth}
  \caption{Cross-study validation results with LightGBM}
  \label{tab:LGBM}
  \ra{1.2}
  \begin{tabular}{cccccc}
    \toprule
    \multirow{2}{*}[-10pt]{Training Set} & \multicolumn{5}{c}{Testing Set}\\
    \cmidrule{2-6}
    & NCI60 & CTRP & GDSC & CCLE & gCSI \\
    \midrule
    NCI60 & \mycell{R$^2$ = 0.74\\MAE = 21.1} & \mygcell{R$^2$ = 0.32\\MAE = 35.7} & \mygcell{R$^2$ = 0.20\\MAE = 36.6} & \mygcell{R$^2$ = 0.40\\MAE = 35.0} & \mygcell{R$^2$ = 0.42\\MAE = 34.6} \\
    CTRP & \mygcell{R$^2$ = 0.42\\MAE = 31.1} & \mycell{R$^2$ = 0.67\\MAE = 23.3} & \mygcell{R$^2$ = 0.25\\MAE = 35.0} & \mygcell{R$^2$ = 0.56\\MAE = 30.0} & \mygcell{R$^2$ = 0.54\\MAE = 30.8} \\
    GDSC & \mygcell{R$^2$ = 0.30\\MAE = 35.5} & \mygcell{R$^2$ = 0.19\\MAE = 38.0} & \mycell{R$^2$ = 0.55\\MAE = 27.3} & \mygcell{R$^2$ = 0.41\\MAE = 36.0} & \mygcell{R$^2$ = 0.58\\MAE = 29.3} \\
    CCLE & \myycell{R$^2$ = 0.05\\MAE = 43.2} & \myycell{R$^2$ = -0.01\\MAE = 44.5} & \myycell{R$^2$ = -0.03\\MAE = 43.3} & \mycell{R$^2$ = 0.60\\MAE = 28.7} & \mygcell{R$^2$ = 0.26\\MAE = 38.5} \\
    gCSI & \mygcell{R$^2$ = 0.19\\MAE = 41.1} & \myycell{R$^2$ = -0.02\\MAE = 45.2} & \myycell{R$^2$ = -0.03\\MAE = 45.1} & \mygcell{R$^2$ = 0.22\\MAE = 41.8} & \mycell{R$^2$ = 0.50\\MAE = 36.7} \\
    \bottomrule
  \end{tabular}
  \end{minipage}
  \end{center}
\end{table*}

\begin{table*}
  \begin{center}
  \begin{minipage}{0.6\textwidth}
  \caption{Cross-study validation results with a multitasking deep learning model}
  \label{tab:UnoMT}
  \ra{1.2}
  \begin{tabular}{cccccc}
    \toprule
    \multirow{2}{*}[-10pt]{Training Set} & \multicolumn{5}{c}{Testing Set}\\
    \cmidrule{2-6}
    & NCI60 & CTRP & GDSC & CCLE & gCSI \\
    \midrule
    NCI60 & \mycell{R$^2$ = 0.81\\MAE = 17.1} & \myGcell{R$^2$ = 0.38\\MAE = 32.2} & \myGcell{R$^2$ = 0.24\\MAE = 35.3} & \myGcell{R$^2$ = 0.48\\MAE = 33.4} & \myGcell{R$^2$ = 0.46\\MAE = 33.4} \\
    CTRP & \myGcell{R$^2$ = 0.44\\MAE = 29.8} & \mycell{R$^2$ = 0.68\\MAE = 22.7} & \mygcell{R$^2$ = 0.23\\MAE = 34.4} & \myGcell{R$^2$ = 0.61\\MAE = 28.3} & \myGcell{R$^2$ = 0.60\\MAE = 28.5} \\
    GDSC & \myGcell{R$^2$ = 0.32\\MAE = 34.0} & \myGcell{R$^2$ = 0.25\\MAE = 36.7} & \mycell{R$^2$ = 0.53\\MAE = 27.2} & \myGcell{R$^2$ = 0.50\\MAE = 32.6} & \myGcell{R$^2$ = 0.60\\MAE = 29.2} \\
    CCLE & \myGcell{R$^2$ = 0.27\\MAE = 36.9} & \myGcell{R$^2$ = 0.20\\MAE = 39.2} & \myGcell{R$^2$ = 0.11\\MAE = 38.9} & \mycell{R$^2$ = 0.68\\MAE = 25.4} & \myGcell{R$^2$ = 0.39\\MAE = 34.2} \\
    gCSI & \myycell{R$^2$ = 0.00\\MAE = 44.9} & \myGcell{R$^2$ = 0.11\\MAE = 43.1} & \myycell{R$^2$ = 0.05\\MAE = 42.8} & \myGcell{R$^2$ = 0.33\\MAE = 40.6} & \mycell{R$^2$ = 0.80\\MAE = 19.2} \\
    \bottomrule
  \end{tabular}
  \end{minipage}
  \end{center}
\end{table*}

As expected, the diagonal cells have relatively high scores for they represent cross validation done on the same data set.
We are more interested in the non-diagonal values because they are less likely to be an artifact of overfitting.
The non-diagonal cells of the matrix are color coded as follows: green for $R^2>0.1$, red for $R^2<-0.1$, and yellow otherwise.
As we mentioned in the data setup, since each cross-study validation experiment involved training multiple models and filling out a matrix of inference results, we limited ourselves to a subset of drugs from NCI60.
For the remaining five studies, all cell lines and drugs were included.
As for input features, we used cell line gene expression profiles, drug chemical descriptors and molecular fingerprints. For details on feature processing, see the Methods section.

Random Forest models trained on CTRP, NCI60 and GDSC achieved moderate cross-study prediction accuracy.
CTRP-trained models performed the best, scoring the highest cross-study $R^2$ of $0.45$ when CCLE was used as the testing set.
CCLE-trained models had less generalization value, and the ones trained on gCSI did not generalize at all.
This was not surprising as gCSI had the smallest number of drugs and thus prone to overfitting.

LightGBM is a gradient boosting version of tree based learning algorithm and is generally considered superior to Random Forest.
Here, the LightGBM models outperformed Random Forest for most of the cells (Table \ref{tab:LGBM}).
However, in the Random Forest matrix, the diagonal values were generally comparable to other cells, suggesting there was little overfitting (with the exception the gCSI row).
In contrast, each diagonal cell in the LightGBM matrix was better than other cells in their row or column.
This was consistent with the view that cross validation within a study was an easier task than cross-study generalization.
Overall, the average improvement in $R^2$ for corresponding cells between Random Forest and LightGBM models was $0.35$ for the diagonal values and $0.19$ for the cross-study comparisons.


\subsubsection{Deep learning performance}
\begin{figure}[htpb]
  \centering
  \includegraphics[width=1.0\linewidth]{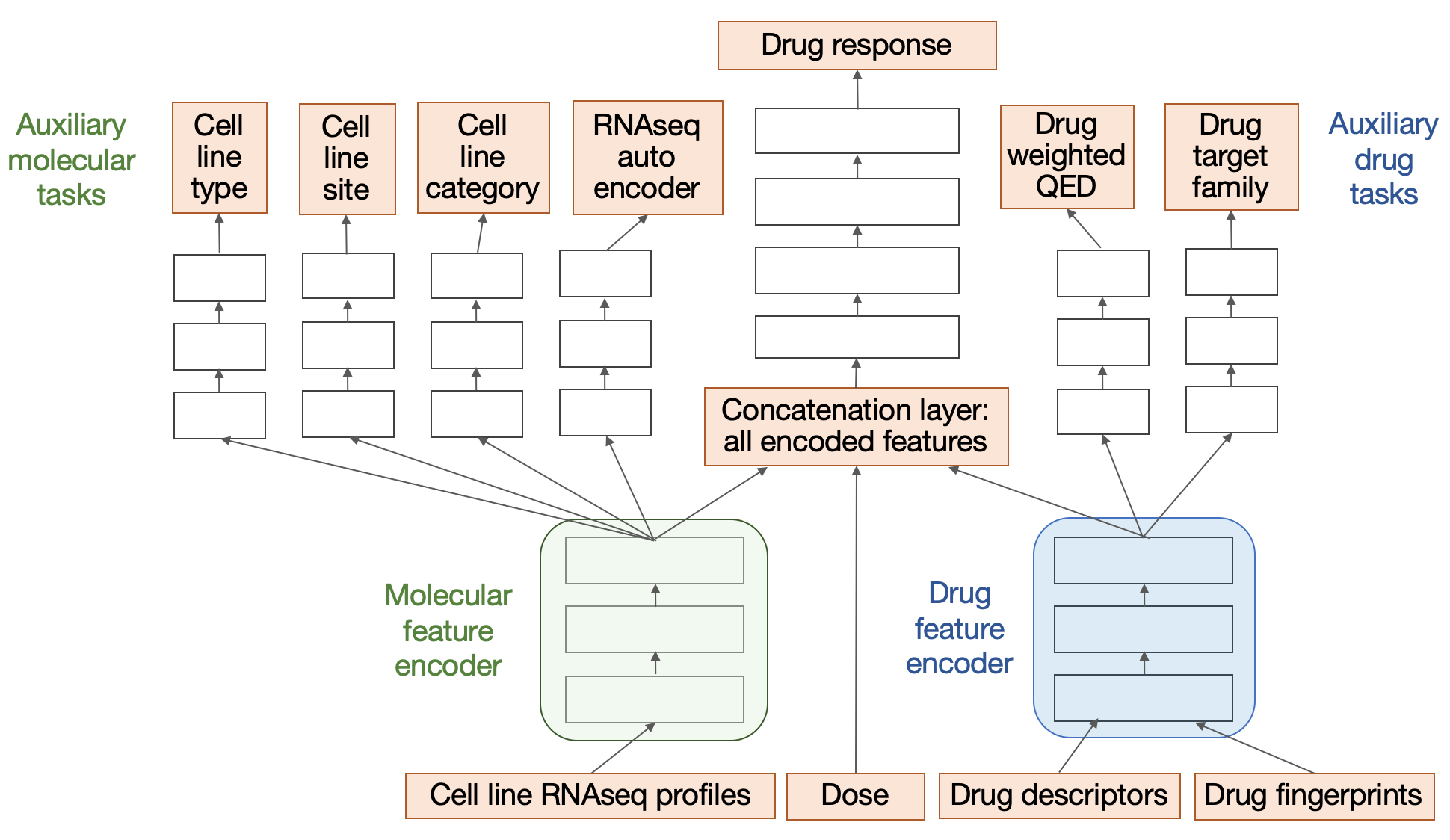}
  \caption{{\bf An example configuration of the multitasking drug response prediction network (UnoMT).} The network predicts a number of cell line properties (tissue category, tumor site, cancer type, gene expression autoencoder) as well as drug properties (target family, drug-likeness score) in addition to drug response.}
  \label{fig:UnoMT}
\end{figure}

Deep learning models generally performed on par or slightly better than LightGBM.
We experimented with a variety of neural network model architectures, and our best prediction accuracy was achieved by a multitasking model that we called UnoMT.
Fig. \ref{fig:UnoMT} shows an example configuration of the UnoMT model where, in addition to the central task of predicting drug response, the model also tried to perform a variety of auxiliary classification and regression tasks (see Methods for details).
These multitasking options allowed the model to use additional labeled data to improve its understanding of cell line and drug properties.

The best performance achieved by deep learning is shown in Table \ref{tab:UnoMT} with three additional prediction tasks turned on: cell line category (normal vs tumor), site, and type.
On average, the cross-study $R^2$ improved $0.11$ over LightGBM models, and the model did nearly perfectly on the additional tasks such as cancer type prediction (not shown).
On models trained on CCLE data, deep \rev{learning} offered the most pronounced improvement.
While the within-study $R^2$ was nearly identical to that of LightGBM, the models were able to, for the first time, generalize to NCI60, CTRP, and GDSC to a moderate degree.
The improved cells in Table \ref{tab:UnoMT} compared to both Tables \ref{tab:RF} and \ref{tab:LGBM} are highlighted in bold.
\rev{Stacking LightGBM and deep learning models resulted in marginal improvement in cross-validation accuracy but did not improve cross-study generalizability.}

\subsubsection{Predictive and predictable cell line data sets}
Used as training sets, each of the five cell line studies was ranked by how predictive they were of the remaining four cell line studies.
Used as testing sets, each of the five cell lines studies were ranked by how predictable they were, using the other four cell line studies as training sets.
By multiple measures (average, median and minimum $R^2$), machine learning algorithms trained on CTRP yield the most accurate predictions on the remaining testing data sets.
This ranking was consistent across the results from all three machine learning models.
By average and median $R^2$ on deep learning results, the gCSI data set was the most predictable cell line dataset, and CCLE was a close second.

\subsection{How model generalizability improves with more data}

\begin{figure*}[htpb]
  \centering
  \includegraphics[width=0.8\linewidth]{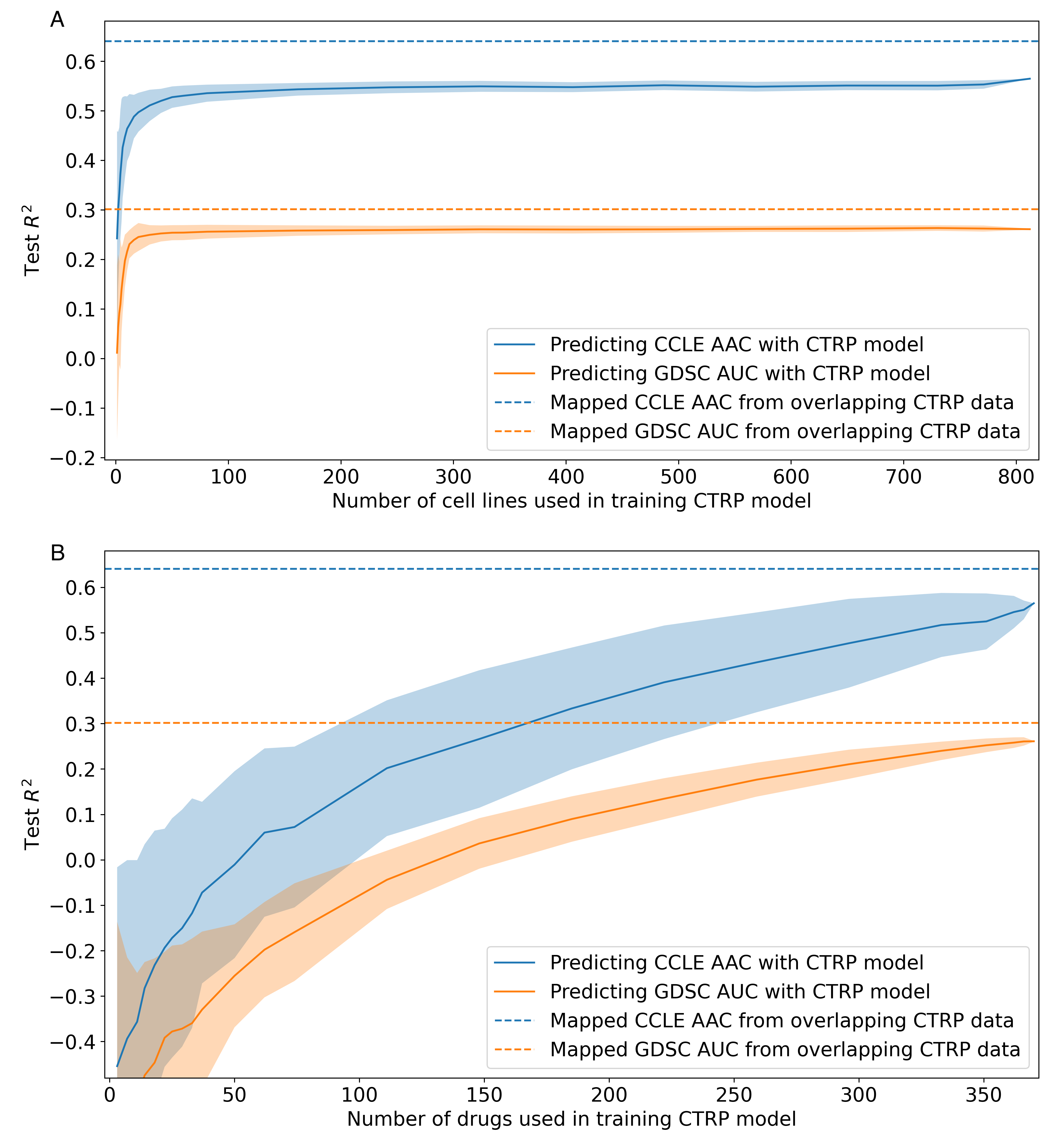}
  \caption{{\bf Impact of cell line or drug diversity on model generalizability.} Models trained on partial CTRP data are tested on CCLE and GDSC. Shades indicate the standard deviation of the cross-study R$^2$. {\bf A}. Models trained with all CTRP drugs and a fraction of cell lines. {\bf B}. Models trained with all CTRP cell lines and a fraction of drugs.}
  \label{fig:frac}
\end{figure*}

Experimental screening data are expensive to acquire.
It is thus critical to understand, from a machine learning perspective, how additional cell line or drug data impact model generalizability.
Data scaling within a single study have been previous explored \cite{partin2020learning}.
In our cross-study machine learning predictions, we observed that the models with poor generalizability tended to have been trained on a small number of drugs (CCLE and gCSI).
To study the relative importance of cell line versus drug diversity, we conducted simulations on CTRP, the most predictive dataset of the five.
We varied the fraction of randomly selected cell lines and kept all the drugs, and vice versa.
The results are plotted in Fig. \ref{fig:frac}.

Models trained on a small fraction of cell lines but all drugs could still quickly approach the full model performance, whereas models trained on limited drug data suffered from low accuracy and high uncertainty.
In either case, it was far more difficult to predict response in a target dataset using a different viability assay (GDSC) than one with the same assay (CCLE).
Inferred upper bounds (dotted lines) were loosely extrapolated from direct mapping results based on data intersection using the best dose-independent metric from Table \ref{tab:var_cross}.

\revbegin
When the numbers of samples were fixed, the model performance was not particularly sensitive to feature selection.
Our previous work showed that deep learning models could still train well with a high dropout ratio of $0.45$, suggesting redundancy in both cell line and drug features \cite{clyde2020systematic}.
Here we also performed data-driven feature selection.
Models trained with the top 1,000 cell line features and top 1,000 drug features, identified based on feature importance scores on a first round of training with LightGBM, had no loss in generalizability.
Models trained with the top 100 cell line features and top 100 drug features reached $90\%$ peak generalizability from CTRP to CCLE and $85\%$ generalizability from CTRP to GDSC.
\revend

\section{Methods}
\label{sec:methods}

\subsection{Feature selection and processing}
Drug response is modeled as a function of cell line features and drug properties.
We used three basic types of features in this study: RNAseq gene expression profiles, drug descriptors, and drug fingerprints.
Drug concentration information was also needed for dose response prediction.
The gene expression values were the batch-corrected RNA-seq measurements from the NCI60, GDSC, and CCLE. Because CTRP and gCSI data sets did not provide RNA-seq data, these data sets were assigned gene expression values from CCLE based on cell line name matching.
\rev{CCLE has molecular characterization data for more than 1,000 cell lines \cite{ghandi2019next}, which cover the CTRP and gCSI cell line sets.}
Log transformation was used to process FPKM-UQ normalized gene expression profiles \cite{shahriyari2019effect}.
Batch correction was performed using a simple linear transformation of the gene expression values such that the per-dataset mean and variance was the same for all data sets \cite{clyde2020systematic}.
To reduce the training time, only the expression values of the LINCS1000 (Library of Integrated Network-based Cellular Signatures) landmark genes \cite{koleti2018data} were used as features.
\revbegin
Our previous work on systematic featurization \cite{clyde2020systematic} found no superset of LINCS1000 (including the full set of over 17,000 genes or known oncogene sets) that clearly outperformed LINCS1000.
\revend
Drug features included 3,838 molecular descriptors and 1,024 path fingerprints generated by the Dragon software package (version 7.0) \cite{dragon7}.
Drug features/fingerprints were computed from 2D molecular structures downloaded from PubChem \cite{kim2019pubchem} and protonated using OpenBabel \cite{o2011open}.

\revbegin
\subsection{Drug response metrics}
Data-driven prediction for drug response is based on the hypothesis that the interaction between tumor and drug treatment can be modeled as a function of three factors: the cancer genomic system, the compound chemical structure, and the drug concentration.
When the first two factors are given, it is further assumed that the rate-limiting step in the killing of cancer cells is the binding of the drug to a target receptor in the cells.

Intuitively, a higher concentration $x$ would lead to a larger fraction of receptors bound to the drug molecules, or a lower fraction $y$ of surviving cancer cells.
At the same time, a fraction $E_\infty$ of the cancer cells are not susceptible to the drug regardless of the drug dose.
This dose response relationship can be worked out based on thermodynamic equilibrium of the bound drug-target complex, and we adopted the three-parameter Hill Slope equation below used in PharmacoDB \cite{smirnov2017pharmacodb}.

\begin{equation}
y(x) = E_\infty + \dfrac{1 - E_\infty}{1 + (\dfrac{x}{EC_{50}})^{HS}}
\end{equation}

$EC_{50}$ is the drug dose at which half of the target receptors are bound (half-maximal response).
Both $EC_{50}$ and $E_\infty$ depend on cancer cell and drug properties.
The Hill Slope coefficient $HS$ quantifies the degree of interaction between binding sites.

With this model, we consistently fit the dose response data from all the studies.
Examples of the resulting dose response curves are shown in Fig. \ref{fig:curves}.
This enabled the derivation of three dose-independent response metrics for comparing across experiments that used different dose levels:

\begin{itemize}
  \item AUC: area under drug response curve for $[10^{-10}, 10^{-4}]{\mu}M$, a fixed dose range.
  \item AAC: area above drug response curve for the measured dose range in a study (same definition in PharmacoDB).
  \item DSS: drug sensitivity score (same to DSS1 in PharmacoDB).
\end{itemize}

DSS is similar to AAC but is potentially more robust as it tries to calibrate the intended dose range of the drug.
Specifically, it aggregates the response over the range where the drug response $R$ exceeds an activity threshold $A_{min}$ \cite{yadav2014quantitative}:
\begin{equation}
DSS \sim \int_{R > A_{min}}{R(x) dx}
\end{equation}

\subsection{Evaluation metrics}

In this study, we chose two commonly used metrics to evaluate the performance of machine learning models.
Mean absolute error (MAE) measures the mean absolute difference between the observed responses $y_i$ and the predicted responses $\hat{y_i}$.
$R^2$, also known as coefficient of determination, measures the explained variance as a proportion of the total variance.

\begin{equation}
MAE = \sum_i{\lvert y_i -\hat{y}_i \rvert}
\end{equation}

\begin{equation}
R^2 = 1 - \dfrac{\sum_i{(y_i -\hat{y}_i)^2}}{\sum_i{(y_i -\overline{y})^2}}
\end{equation}

$R^2$ was additionally used to evaluate the response variability levels within and across studies.
This was done by examining the observed response values on the overlapping drug and cell line groups from different experimental replicates or studies.
For example, a high $R^2$ score would suggest good agreement among the multiple measurements and thus low variability in experimental data.
\revend

\subsection{Machine learning}
Three different machine learning algorithms were evaluated in this study: Random Forest, LightGBM and deep neural networks.
The cell line prediction error for all methods was assessed using the mean absolute error (MAE) and Scikit-learn \cite{scikit-learn} definition of $R^2$ value, which is equal to the fractional improvement in mean squared error (MSE) of the method compared to the MSE of the prediction method that outputs the average response of the test set, independent of dose, gene expression and drug features.
In the diagonal cells in the matrices (Tables \ref{tab:RF}, \ref{tab:LGBM}, and \ref{tab:UnoMT}), mean values of 5-fold cross validation partitioned on cell lines are reported.
For cross-study analysis, the entire selected set of source study data were used to train the model, and the non-diagonal cells report test metrics on the whole target data set.
The Random Forest models were trained using the default Scikit-learn implementation (version 0.22).
The LightGBM models were trained using the Scikit-learn implementation with the number of trees set to be proportional to the number of drugs included in the training set.

\subsection{Deep learning}
The reported deep neural network is based on a unified model architecture, termed Uno, for single and paired drug treatment.
This architecture was extended from a previously developed neural network called ``Combo'' \cite{xia2018predicting} to simultaneously optimize for feature encoding and response prediction.
Cell line and drug features first pass through their separate feature encoder subnetworks before being concatenated with drug concentration to predict dose response through a second level of subnetwork.
In the multitasking configuration of the network (UnoMT), the output layers of molecular and drug feature encoders are further connected with additional subnetworks to predict auxiliary properties.
All subnetworks have residual connections between neighboring layers \cite{he2016deep}.
The multitasking targets include drug-likeness score (QED \cite{bickerton2012quantifying}) and target family for drugs, and tissue category (normal vs. tumor), type, site, and gene expression profile reconstruction for molecular samples.
Not all labels were available for all samples.
In particular, tissue category and site applied only to the Cancer Genome Atlas (TCGA \cite{tomczak2015cancer}) samples but not the cell line studies.
We included them, nonetheless, to boost the model's understanding of gene expression profiles (data downloaded from the NCI's Genomic Data Commons \cite{jensen2017nci}).
The drug target information was obtained through ID mapping and SMILES string search in ChEMBL \cite{gaulton2012chembl}, BindingDB \cite{gilson2016bindingdb}, and PubChem databases.
The binding targets curated by these sites were then mapped to the PANTHER gene ontology \cite{mi2013large}.
326 out of the 1,801 combined compounds had known targets for the exact SMILES match.
A Python data generator was used to join the cell line and drug features, the response value, and other labels on the fly.
For multitasking learning, the multiple partially labeled data collections were trained jointly in a round-robin fashion for each target, similar to how generative adversarial networks (GANs) \cite{goodfellow2014generative} take turns to optimize loss functions for generators and discriminators.

\section{Discussion}
In this study, we sought to understand the performance of drug response prediction models in a broader context.
We hypothesized that the observed $R^2$ score barrier of $0.8$ \cite{baptista2021deep} might partly be a result of variability in drug response assay.
Indeed, we found that the measured dose response values differed considerably among replicates in a single study.
In the case of GDSC, this variability in terms of standard deviation was more than $10\%$ of the entire drug response range.
Practically, this meant that, if we used one response value as prediction for another from the same cell-drug-dose combination, we would only obtain an average $R^2$ of $0.81$.
The cause for this variability is not well understood, but experimental protocol likely plays a big role, as evidenced by the lower variability observed among NCI60 replicates.
Standardization in experimental design will be key to maximizing the value of screening data for joint learning.

When we moved beyond single studies to compare drug response values across studies, the variability increased.
This phenomenon has been discussed by numerous previous studies \cite{haibe2013inconsistency,mpindi2016consistency,safikhani2016revisiting,sadacca2017new} from a statistical consistency perspective.
In this study, we approached it from a machine learning angle.
Using the best available integrative metrics, we compared dose-independent response values across different studies.
We arrived at rough upper bounds for how well models trained on one data source could be expected to perform in another.
These numbers depended on whether the source and target studies used the same cell viability assay.
In the case of identical assay, the $R^2$ ceiling was estimated to be $0.65$ between CTRP and CCLE.
In the case of different assays, it was markedly lower, at $0.31$, between CTRP and GDSC.

These estimates put the recent progress in machine learning based drug response prediction into perspective.
We suggest that cross-study evaluation be used as an additional tool for benchmarking model performance; without it, it's difficult to know how much of the model improvement is generalizable.
We illustrated this point with systematic characterization of cross-study performance using three different machine learning approaches.
For example, going from a simple Random Forest model to LightGBM trained on CTRP, accuracy improved over $220\%$ judging by cross validation $R^2$.
However, the improvement on model generalization to CCLE was only $24\%$.
In some cases, extrapolation error actually increased as within-study performance improved.
For an opposite example, the deep learning models made only marginal improvement over LightGBM in within-study performance, but the cross-study improvement, averaged $0.11$ in $R^2$, was much more appreciable.
This may be somewhat counterintuitive since neural networks are known to be overparameterized and prone to overfitting.
However, as we have demonstrated with a multitasking model, the high capacity of deep learning models could be put to work with additional labeled data in auxiliary tasks.

Drug screening experiments are costly.
How should we prioritize the acquisition of new data?
A recent study showed with leave-one-out experiments that drug response models had much higher error when extrapolating to new drugs than new cell lines \cite{li2019deepdsc}.
This is consistent with our finding.
The models that did not generalize well tended to have been trained on data sets with fewer drugs (CCLE and gCSI).
Further, when we withheld samples from training drug response models, we found that the loss of drug data was significantly more crippling than that of cell lines.

\revbegin
In addition to increasing the number of screened drugs, a careful selection based on mechanistic understanding or early experimental data may also help.
As we have seen in \Cref{tab:RF,tab:LGBM,tab:UnoMT}, models trained on CTRP generalize well and this was not limited to studies with the same viability assay.
A good example is the NCI60 column: as the target study, NCI60 used a different viability assay from both CTRP and GDSC, yet CTRP's prediction accuracy was notably higher than GDSC's.
This may have to do with CTRP's design of an \emph{Informer Set} of 481 compounds that target over 250 diverse proteins, covering a wide range of cell processes linked to cancer cell line growth \cite{seashore2015harnessing}.
Some probe molecules had no known protein targets but were selected for their ability to elicit distinct changes in gene expression profiles.
\revend
Taken together, this suggests that it would be beneficial for future screening studies to prioritize drug diversity.
Given the vast design space of potentially active chemical compounds, estimated to be in the order of $10^{60}$ \cite{bohacek1996art}, intelligent methods that can reason about molecule classes are needed.

\section{Conclusion}
Precision oncology requires precision data.
In this article, we reviewed five cancer cell line screening data sets, with a focus on drug response variabilities within and across studies.
We demonstrated that these variabilities put constraints on the performance of machine learning models, in a way not obvious to traditional cross validation within a single study.
Through systematic analysis, we compared how models trained on one data set extrapolated to another, revealing a wide range in predictive power of both study data and machine learning methods.
While deep learning results are promising, future success of drug response prediction will rely on the improvement of model generalizability.
Experimental standardization and prioritization in drug diversity will also be key to maximizing the value of screening data for integrative learning.
\revbegin
The integrated data files from this study are available in the Predictive Oncology Model \& Data Clearinghouse hosted at the National Cancer
Institute\footnote{\rev{\url{https://modac.cancer.gov/assetDetails?dme_data_id=NCI-DME-MS01-8088592}}}.
\revend

\vspace{2em}
\fbox{
\begin{minipage}{0.8\linewidth}
\section{Key Points}
\begin{itemize}
\item Cross validation in a single study overestimates the accuracy of drug response prediction models, and differences in response assay can limit model generalizability across studies.
\item Different machine learning models have varying performance in cross-study generalization, but they generally agree that models trained on CTRP are the most predictive and gCSI is the most predictable.
\item Based on simulated experiments, drug diversity, more than tumor diversity, is crucial for raising model generalizability in preclinical screening.
\end{itemize}
\end{minipage}
}




\section{Acknowledgments}
We thank Marian Anghel for helpful discussions.
This work has been supported in part by the Joint Design of Advanced Computing Solutions for Cancer (JDACS4C) program established by the U.S. Department of Energy (DOE) and the National Cancer Institute (NCI) of the National Institutes of Health. This work was performed under the auspices of the U.S. Department of Energy by Argonne National Laboratory under Contract DE-AC02-06-CH11357, Lawrence Livermore National Laboratory under Contract DE-AC52-07NA27344, Los Alamos National Laboratory under Contract DE-AC5206NA25396, and Oak Ridge National Laboratory under Contract DE-AC05-00OR22725. This project has also been funded in whole or in part with federal funds from the National Cancer Institute, National Institutes of Health, under Contract No. HHSN261200800001E. The content of this publication does not necessarily reflect the views or policies of the Department of Health and Human Services, nor does mention of trade names, commercial products, or organizations imply endorsement by the U.S. Government.



\setcitestyle{numbers}
\bibliographystyle{unsrt}
\bibliography{BIB-draft}



\vspace{8em}
\begin{biography}{}
  {\author{Fangfang Xia} is a Computer Scientist in Data Science and Learning Division at Argonne National Laboratory.
    \author{Jonathan Allen} is a Bioinformatics Scientist at Lawrence Livermore National Laboratory.
    \author{Marian Anghel} is a Technical Staff Member at Los Alamos National Laboratory.
    \author{Prasanna Balaprakash} is a Computer Scientist in Mathematics and Computer Science Division at Argonne National Laboratory.
    \author{Thomas Brettin} is a Strategic Program Manager in Computing, Environment and Life Sciences at Argonne National Laboratory.
    \author{Cristina Garcia-Cardona} is a Scientist at Los Alamos National Laboratory.
    \author{Austin Clyde} is a Computational Scientist at Argonne National Laboratory and a PhD student in Computer Science at University of Chicago.
    \author{Judith Cohn} is a Research Scientist at Los Alamos National Laboratory.
    \author{James Doroshow} is the Director of Division of Cancer Treatment and Diagnosis at National Cancer Institute.
    \author{Xiaotian Duan} is a PhD student in Computer Science at University of Chicago.
    \author{Veronika Dubinkina} is a PhD student in Department of Bioengineering at University of Illinois at Urbana-Champaign.
    \author{Yvonne Evrard} is an Operations and Program Manager at Frederick National Laboratory for Cancer Research.
    \author{Ya Ju Fan} is a Computational Scientist in Center for Applied Scientific Computing at Lawrence Livermore National Laboratory.
    \author{Jason Gans} is a Technical Staff Member at Los Alamos National Laboratory.
    \author{Stewart He} is a Data Scientist at Lawrence Livermore National Laboratory.
    \author{Pinyi Lu} is a Bioinformatics Analyst at Frederick National Laboratory for Cancer Research.
    \author{Sergei Maslov} is a Professor and Bliss Faculty Scholar in Department of Bioengineering and Physics at the University of Illinois at Urbana-Champaign and holds joint appointment at Argonne National Laboratory.
    \author{Alexander Partin} is a Computational Scientist in Data Science and Learning Division at Argonne National Laboratory.
    \author{Maulik Shukla} is a Project Lead and Computer Scientist in Data Science and Learning Division at Argonne National Laboratory.
    \author{Eric Stahlberg} is the Director of Biomedical Informatics and Data Science at the Frederick National Laboratory for Cancer Research.
    \author{Justin M. Wozniak} is a Computer Scientist in Data Science and Learning Division at Argonne National Laboratory.
    \author{Hyunseung Yoo} is a Software Engineer in Data Science and Learning Division at Argonne National Laboratory.
    \author{George Zaki} is a Bioinformatics Manager at Frederick National Laboratory for Cancer Research.
    \author{Yitan Zhu} a Computational Scientist in Data Science and Learning Division at Argonne National Laboratory.
    \author{Rick Stevens} is an Associate Laboratory Director at Argonne National Laboratory and a Professor in Computer Science at University of Chicago.
  }
\end{biography}


\end{document}